\documentstyle[11pt,amssymb,epsfig,amsmath]{article}

\begin{document}
\newcommand{\be}{\begin{equation}}
\newcommand{\ee}{\end{equation}}
\newcommand{\ba}{\begin{eqnarray}}
\newcommand{\ea}{\end{eqnarray}}
\newcommand{\no}{\nonumber \\}
\newcommand{\gsim}{\mathrel{\hbox{\rlap{\lower.55ex \hbox {$\sim$}}
                   \kern-.3em \raise.4ex \hbox{$>$}}}}
\newcommand{\lsim}{\mathrel{\hbox{\rlap{\lower.55ex \hbox {$\sim$}}
                   \kern-.3em \raise.4ex \hbox{$<$}}}}

\def\be{\begin{eqnarray}}
\def\ee{\end{eqnarray}}
\def\bea{\be}
\def\eea{\ee}
\newcommand{\e}{{\mbox{e}}}
\def\del{\partial}
\def\vr{{\vec r}}
\def\vk{{\vec k}}
\def\vq{{\vec q}}
\def\vp{{\vec p}}
\def\vP{{\vec P}}
\def\vt{{\vec \tau}}
\def\vs{{\vec \sigma}}
\def\vJ{{\vec J}}
\def\vB{{\vec B}}
\def\hatr{{\hat r}}
\def\hatk{{\hat k}}
\def\roughly#1{\mathrel{\raise.3ex\hbox{$#1$\kern-.75em%
\lower1ex\hbox{$\sim$}}}}
\def\lsim{\roughly<}
\def\gsim{\roughly>}
\def\fm{{\mbox{fm}}}
\def\vx{{\vec x}}
\def\vy{{\vec y}}
\def\({\left(}
\def\){\right)}
\def\[{\left[}
\def\]{\right]}
\def\EM{{\rm EM}}
\def\barp{{\bar p}}
\def\zz{{z \bar z}}
\def\mus{{\cal M}_s}
\def\abs#1{{\left| #1 \right|}}
\def\ve{{\vec \epsilon}}
\def\nlo#1{{\mbox{N$^{#1}$LO}}}
\def\MS{{\mbox{M1V}}}
\def\mut{{\mbox{M1S}}}
\def\Qt{{\mbox{E2S}}}
\def\rM{{\cal R}_{\rm M1}}\def\rE{{\cal R}_{\rm E2}}
\def\la{{\Big<}}
\def\ra{{\Big>}}
\def\lsim{\mathrel{\rlap{\lower3pt\hbox{\hskip1pt$\sim$}}
     \raise1pt\hbox{$<$}}} %less than or approx. symbol
\def\gsim{\mathrel{\rlap{\lower3pt\hbox{\hskip1pt$\sim$}}
     \raise1pt\hbox{$>$}}} %greater than or approx. symbol
\def\N{${\cal N}\,\,$}

\def\ka{\kappa}
\def\lam{\lambda}
\def\dlt{\delta}
\def\eps{\epsilon}
\def\omg{\omega}
\def\lv{\lvert}
\def\rv{\rvert}
\def\xp{x^+}
\def\xm{x^-}
\def\xperp{x_{\perp}}
\def\bph{\bar{\phi}}

\def\J#1#2#3#4{ {#1} {\bf #2} (#4) {#3}. }
\def\PRL{Phys. Rev. Lett.}
\def\PL{Phys. Lett.}
\def\PLB{Phys. Lett. B}
\def\NP{Nucl. Phys.}
\def\NPA{Nucl. Phys. A}
\def\NPB{Nucl. Phys. B}
\def\PR{Phys. Rev.}
\def\PRC{Phys. Rev. C}

\renewcommand{\thefootnote}{\arabic{footnote}}
\setcounter{footnote}{0}

\vskip 0.4cm \hfill { }
 \hfill {\today} \vskip 1cm

\begin{center}
{\LARGE\bf On the critical condition in gravitational 
shock wave collision and heavy ion collisions
%Shock wave collisions in $AdS_5$ and heavy ion collisions
%  from AdS/CFT duality
   }
\date{\today}

\vskip 1cm {\large Shu
Lin${}^a$\footnote{E-mail:slin@mppmu.mpg.de},
and Edward
Shuryak${}^b$\footnote{E-mail:shuryak@tonic.physics.sunysb.edu}
 }

%{\large and}

\end{center}

\vskip 0.5cm

\begin{center}

 {
${}^a$\it Max-Planck-Institut f\"{u}r Physik (Werner-Heisenberg-Institut) \\ F\"{o}hringer Ring 6, 80805 M\"{u}nchen, Germany\\
${}^b$\it Department of Physics and Astronomy, SUNY Stony-Brook, NY 11794}

\end{center}

\vskip 0.5cm

\begin{abstract}

In this paper, we derived a critical condition for matter 
equilibration in heavy ion
collisions using a holographic approach. A gravitational shock waves with 
infinite transverse extension 
is used to model infinite nucleus. 
We constructed the trapped surface in the collision of 
two asymmetric planar shock waves with sources at different depth in the bulk
AdS and formulated a critical condition for 
matter equilibration in collision of ``nucleus'' in the dual gauge theory. 
We found the critical condition is insensitive to the depth of the source
closer to the AdS boundary.
To understand the origin of the critical condition, we computed the
 Next to Leading
Order stress tensor in the boundary field theory due to the interaction
of the nucleus and found the critical condition corresponds 
to the breaking down of the perturbative expansion. We indeed expect non-perturbative effects be needed to describe black hole formation.
\end{abstract}

%\preprint{MPP-2010-144}

\newpage

\renewcommand{\thefootnote}{\#\arabic{footnote}}
\setcounter{footnote}{0}

\section{Introduction}

The AdS/CFT correspondence is conjectured as a duality between weakly
coupled gravity theory and strongly coupled $\cal N$=4 Super Yang-Mills theory in the
limit of large $N_c$ and strong coupling\cite{maldacena,gubser,witten}. Its
applications to strongly coupled Quark Gluon Plasma (sQGP) have revealed many
novel features of the strongly coupled medium such as very low viscosity\cite{KSS}, absence of jets, Mach cone formation
and other hydrodynamical phenomena, see e.g. review \cite{Shuryak:2008eq}. While static and
near equilibrium properties have been extensively studied in this context, deriving corrections to 
hydrodynamics, the out-of-equilibrium  aspects of the strongly coupled gauge theory remains
less understood. One of the main challenges in heavy ion collisions remains the
understanding of early equilibration of matter produced in the collisions.

 Recently there
have been several attempts to model the initial non-equilibrium stage of the collision, including
 \cite{weak} and \cite{chesler}. Our paper \cite{Lin:2008rw}  provided relatively simple description of black hole formation,
 due to elastic membrane falling under its own weight. In this case the ultimate equilibration is always assured, and it happens as a gradual
 propagation of the equilibration boundary in the scale space (along the 5-th holographic dimension), from the ultraviolet
 (UV) toward the infrared (IR) direction. 

%This would involve knowledge of strongly coupled gauge theory in far from
%equilibrium regime.
 According to principles of AdS/CFT,  due to large $N_c$ limit all issues have to be understood in terms of classical gravity problem.  
Thermal equilibration of
matter and early entropy production is in this setting 
 dual to the formation of a (black hole) horizon, trapping some amount of information 
from the distant observer, where our world is. This mechanism not only is able to
provide some lower bound on the amount of {\em entropy production} in the collision, but it also provides qualitative ``yes'' or ``no'' answer
if the information trapping does or does not happen,
as a function of given initial condition  of the problem. Thus one of the interesting unexpected features of the 
problem are some rapid transition into a new regime, 
as a function of e.g. collision energy, density of the colliding objects or (not discussed in this work)
the impact parameter of the collisions.

The relation between the trapped surface at the collision moment and the lower bound on the entropy production 
has been introduced by Gubser, Pufu and  Yarom \cite{GPY}, who have considered collision of ultrarelativistic small black holes in AdS$_5$.
It can be viewed as 
 a collision of gravitational shock waves, having near-zero longitudinal width but possessing a certain profile in  3 transverse
 coordinates $x^2,x^3$ and the holographic coordinate $z$. Mathematically, the trapped surface at the collision point satisfies the Laplace eqn, plus certain nontrivial boundary conditions on the surface.
% In their paper the collision itself has not been folllowed: instead
% they looked at 
 % the formation of trapped
%surface at time zero, when the shocks first met. If the trapped surface is formed, it is interpreted as the signal of some matter equilibration.
Furthermore, if a solution to those conditions is found, the trapped surface  area  gives
(the lower bound to) the entropy production in the collisions. Technically
construction of the trapped surface closely follows early works in flat 
space background\cite{penrose,eardley,yoshino,veneziano}. The specific problem addressed in that work \cite{GPY} was
 central collision of two point black holes. Among its important conclusions was 
   e.g. a prediction of the entropy dependence on the (CM) collision energy $S(E) \sim E^{2/3}$.

 This approach has been then generalized
to the non-central collisions. We  found \cite{LS} that  trapped surface formation is not possible
beyond certain  critical impact parameter, depending on the collision energy.  Furthermore, the disappearance of the
trapped surface happens suddenly, as a 1-st order transition. An intriguing observation,  also pointed out in our paper \cite{LS},
is that phenomenologically the multiplicity of the produced particles (the entropy) per participant nucleon
in ultrarelativistic collisions at RHIC also changes rapidly between ``non-thermal'' peripheral and ``thermal'' more central collisions.
   The specific results about the trapped surface
were later confirmed in \cite{GPY2,mozo}. Like it has been the case in flat space, the value of the critical impact parameter can be
understood as a bound on the angular momentum for the shock wave pair
at a given center of mass to form a AdS-Kerr black hole. %It gives the
%correct asymptotic power law energy dependence of the critical impact parameter.

All the above-mentioned works have been using a shock wave arising from a point source
in the bulk (small black holes).  The size of the colliding nuclei were thus incorporated via the distance of those objects from the boundary along the
holographic coordinate $z$.   However, as emphasized in \cite{LS}, this is an oversimplification of the problem.
The transverse extension of the colliding objects in $x^2,x^3$ can be introduced independently of the profile in the holographic $z$ 
direction. The latter, due to very basic features of the AdS/CFT correspondence, should be
 ascribed  instead to  the {\em intrinsic scale} variable, in the sense of the renormalization group,
 describing its microscopic structure. In the collision of ordinary objects it would be interatomic scale,
 for high energy QCD   
 the holographic coordinate $z$ at which the colliding object are before the collisions 
 should represent  the typical scale of their wave function, known as the  {\em ``saturation scale''}.
This scale affects the typical ``equilibration time''
and other properties of the problem, thus should be taken into account.
Other scales can also be important for the equilibration process, e.g. the
longitudinal width of the nucleus was studied in a more recently 
paper\cite{CY}, which
contains a numerical evolution of the nonlinear Einstein equation and found interesting behavior
of slowing down of the nucleus after collision.

In this paper, we will focus on the effect of the saturation scale and
 model it with the simplest possible geometry, proposed for this
purpose in \cite{LS}. It is a collision of {\em  wall} shock waves, which are infinite and homogeneous  in 2 transverse spatial dimensions.
The extension in $z$ of the trapped region  has been found for the
collision of such wall shock waves.
It has been done for the simplest case of a symmetric collision, in which both colliding walls are the same.
We start in this paper discussing a more general case, in which two colliding walls are not the same. Physically, one may
think of two colliding objects  made of different materials with different
densities, which are modeled  by their different ``saturation scales'' 
$\frac{1}{z_1},\frac{1}{z_2}$. The question we will answer  is the precise 
 critical condition on their values $z_1,z_2$  beyond which the trapped surface is not formed.
 
 Perhaps the reader may wander why are we interested in such a question. It is clear that one of the
 most important variable is the energy (rapidity) of the colliding objects: the black holes can only be formed if
 it is large enough. However, let us also
remind the reader that in heavy ion  collisions the energy per nucleon is not the only important variable: for example
rapid equilibration and hydrodynamical behavior experimentally observed at RHIC for collisions of two heavy ions such as AuAu,
 are indeed $not$ observed say for deuteron-Au collisions at the same rapidity of the colliding nuclei. Similarly, we find that two
   walls, made of sufficiently different materials, can also collide $without$ classical equilibration and entropy formation,   
 at the same energy at which the symmetric walls would produce the trapped surface.

Another issue, to be addressed in section 4, deals with the difficult problem of finding the gravitational solution for the non-zero time, in the future quadrant
of the time-longitudinal coordinates. In flat Minkowski space-time this is a long-standing problem of the general relativity. Recent numerical studies\cite{pretorius} have managed to reach gamma factor
of the order of few units and reasonable agreement is observed with previous
partial analytic results reported in \cite{PD,eardley,yoshino,veneziano}. However,  
the problem
 gets even more complicated in the curved 5-dimensional space $AdS_5$ needed for current applications,
see e.g. \cite{romatschke,AKT,taliotis,CY}. 
The ``Next-to-Leading Order'' (NLO) effect we will discuss are the  ``debris'' produced in the shock wave collision, the gravitons radiated perturbatively.
We will compute the NLO correction to the metric, and read the corresponding stress tensor
on the dual field theory on the boundary: such ``early time'' stress tensor plays an important role in the theory of heavy ion collisions, as it provides
the initial conditions for the standard hydrodynamical treatment.
 We will follow most closely the work by Taliotis\cite{taliotis} in the settings:   
the main difference is that our source is localized in the holographic direction $z$, instead of
the transverse directions as in \cite{taliotis}.
%which significantly simplify the overall geometry of the collision. 

%collisions will be the main purpose of this work.

\section{Wall on wall shock wave collision}

We start with the wall shock wave model proposed in \cite{LS}. The metric of
a single shock wave moving in direction $\xp$ is given by:

\be\label{wall}
ds^2=L^2\frac{-d\xp d\xm+d\xperp^2+\phi(z,\xp)d\xp{}^2+dz^2}{z^2}
\ee

The shock wave profile $\phi(z)$ satisfies the following equation:

\be\label{wall_source}
(\del_z^2-\frac{3}{z}\del_z)\phi(z,\xp)=-16\pi G_5\mu\frac{z_0^3}{L^3}\dlt(\xp)\dlt(z-z_0)
\ee

with RHS being the source, which has infinite extension in the directions of
$\xperp$, thus the name wall shock wave.
 The parameter $z_0$ is interpreted as the inverse saturation scale. 
The solution to (\ref{wall_source}) is given by:

\be\label{phi}
\phi(z,\xp)=4\pi G_5\mu\frac{z_0^4}{L^3}\dlt(\xp)
\left\{\begin{array}{l@{\quad}l}
\frac{z^4}{z_0^4}& z\le z_0 \\
1& z>z_0
\end{array}
\right.
\ee

The stress tensor follows from (\ref{phi}) reads:

\be
T_{++}=\mu\dlt(\xp)
\ee

Now consider the collision of two shock waves, as a model of heavy ion
collisions. The metric of the shock waves before collision is given by:

\be\label{phi12}
ds^2=L^2\frac{-d\xp d\xm+d\xperp^2+\phi_1(\xp,z)d\xp{}^2+\phi_2(\xm,z)d\xm{}^2+dz^2}{z^2}
\ee

The shock wave profiles solve the following equations:

\be\label{norm}
&&(\del_z^2-\frac{3}{z}\del_z)\phi_1(\xp,z)=-16\pi G_5\mu_1\frac{z_1^3}{L^3}\dlt(\xp)\dlt(z-z_1) \\
&&(\del_z^2-\frac{3}{z}\del_z)\phi_2(\xp,z)=-16\pi G_5\mu_2\frac{z_2^3}{L^3}\dlt(\xm)\dlt(z-z_2)
\ee

Note we have absorbed the delta function into the definition of the shock
wave profiles as in \cite{taliotis}. The dual stress tensor reads:

\be\label{LO_Tmn}
&&T_{++}=\mu_1\dlt(\xp) \no
&&T_{--}=\mu_2\dlt(\xm)
\ee

The superposition of two shock waves (\ref{phi12}), 
solves the Einstein equation in the region with $\theta(\xp)\theta(\xm)=0$. (Here
$\theta(x)$ is the Heaviside step function, 1 for positive and 0 for negative argument.)
The shock waves only interact
and modifies the metric in the future quadrant $\theta(\xp)\theta(\xm)>0$.
 
As explained in the Introduction, our colliding walls 
are dual to ``nuclei'' of infinite size, 
so the concept of impact 
parameter does not exist. Instead, we have also chosen two nucleus to have
the same energy $\mu_1=\mu_2$, but different saturation scales $z_1\ne z_2$. (To be specific, we demand
$z_1>z_2$.) Although our shock waves are sourced by the delta functions, they have finite size in the $z$ direction,
decreasing both into the UV and the IR. 
%This last point is very important, it separates our approach from the earlier works
This is different from other approaches using sourceless shock waves\cite{JP,romatschke,AKT,beuf,iancu,CY}.

 These finite extension of the shock waves in $z$ explains why the
trapped surface can be found also in a finite interval in $z$, 
 we will call upper and lower positions of the trapped surface 
$z_a,z_b$.
%E.g. non-central collision of black holes
 %in flat space is usually characterised by ``inspiral'', ``merger'' and 
%``ring down'' 
%phases in the
%formation of black hole\cite{pretorius}}. 
%
%Therefore, a perturation in the amplitude of 
%shock wave is valid in early time after the collision. It is interesting to
%see how the perturbative correction to the metric due to the
%shock wave interaction depends on the saturation scales and its implication
%to heavy ion collisions.

The entropy lower bound, dual to the ``area'' of the trapped surface
 is given by:

\be
&&S=\frac{2A}{4G_5}=\frac{\int\sqrt{g}dzd^2x_\perp}{2G_5} \no
&&s\equiv\frac{S}{\int d^2x_\perp}=\frac{L^3}{4G_5}(\frac{1}{z_a^2}-\frac{1}{z_b^2})
\ee

\section{Critical condition for trapped surface formation}

In this section, we will construct the trapped surface associated with the collision of two shock waves. 
Let us for the completeness  recall  the mathematical basis defining the trapped surface. The equations are produced by
required vanishing of the so called ``expansion''
combination: loosely speaking it means that the geodesics of 
forward moving, outgoing massless particles should converge on this surface. 
The limiting case when the geodesics neither converge nor diverge defines the
maginally trapped surface.
 It can be shown to correspond to a relatively simple problem a la electrostatic solution in a cavity (the Laplacian
 with given sources)  with zero boundary condition on the surface,
complemented by additional nontrivial condition for the magnitude of the field derivatives at the surface itself.
Following \cite{GPY,LS,GPY2}, the master equation 
for trapped surface is given by:

\be\label{master}
&&z^2\Psi_i''-z\Psi_i'-3\Psi_i=-16\pi G_5\mu_i z_i^4\dlt(z-z_i) \no
&&\Psi_i(z_a)=\Psi_i(z_b)=0 \no
&&\Psi_1'(z_a)\Psi_2'(z_a)\frac{z_a^2}{L^2}=\Psi_1'(z_b)\Psi_2'(z_b)\frac{z_b^2}{L^2}=4
\ee
with $i=1,2$.
The trapped surface for wall-on-wall shock wave collision is just $z_a<z<z_b$.
The first two equations can be solved as:

\be
&&\Psi_i(z)=\left\{\begin{array}{l@{\quad}l}
C_i\(\frac{z^3}{z_a^3}-\frac{z_a}{z}\)& z<z_i \\
D_i\(\frac{z^3}{z_b^3}-\frac{z_b}{z}\)& z>z_i
\end{array}
\right. \\
&&\text{with} \no
&&C_i=-4\pi G_5\mu_i\frac{(\frac{z_i^4}{z_b^4}-1)z_b}{\frac{z_b^4-z_a^4}{z_a^3z_b^3}} \\
&&D_i=-4\pi G_5\mu_i\frac{(\frac{z_i^4}{z_a^4}-1)z_a}{\frac{z_b^4-z_a^4}{z_a^3z_b^3}}
\ee

We can always apply a longitudinal boost such that both shock waves have
the same energy density $\sqrt{\mu_1\mu_2}$. Then the third equation in (\ref{master}) leads
to
%The existence of the trapped surface is equivalent to the following condition:

\be\label{constraint}
C_1C_2=D_1D_2=\frac{L^2}{4}
\ee

Let us consider the case $z_1=z_2\equiv z_0$ first. (\ref{constraint}) leads to:

\be\label{cardano}
&&z_a+z_b=\frac{8\pi G_5\sqrt{E_1E_2}}{L} =A_1 \\
&&\frac{(z_a+z_b)^2-3z_az_b}{(z_az_b)^3}=\frac{L^3}{z_0^4}={1 \over A_2}
\ee
in which two appearing combinations of parameters are for brevity called $A_1,A_2$. 
The resulting cubic eqn 
\be (z_az_b)^3 + 3 A_2 (z_az_b)-A_1A_2=0  \ee
 can be solved by
Cardano formula. The explicit solution is not illustrative and is not
showed here. We note, however the solution has to satisfy the inequality
$4z_az_b\le(z_a+z_b)^2=A_1^2$, which gives rise to the following constraint:

\be\label{critical_z0}
\frac{2\pi^2}{N_c^2}\mu z_0^3\ge 1
\ee
where we have used $G_5=\frac{\pi L^3}{2N_c^2}$. (\ref{critical_z0}) is the
critical condition for trapped surface formation in a symmetric collision
of gravitational shock waves.

\begin{figure}[t]
\includegraphics[width=0.5\textwidth]{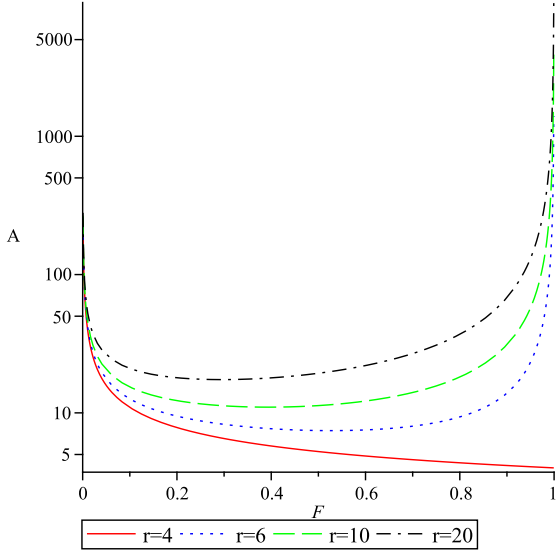}
\caption{\label{ArF}$A$ as a function of $F$. A minimum always
exists in $0<F<1$ for $r>4$. In the extreme case $r=4$, the minimum locates
at $F=1$}
\end{figure}

When $z_1>z_2$, we
define $\frac{z_1^4}{z_a^2z_b^2}=\lam_1,\,\frac{z_2^4}{z_a^2z_b^2}=\lam_2$,
(\ref{constraint}) can be simplified to:
\be\label{lam}
\left\{\begin{array}{l}
\(\frac{z_a}{z_b}\)^2+\(\frac{z_b}{z_a}\)^2+1=\frac{\lam_1+\lam_2+1}{\lam_1\lam_2}\\
(z_az_b)^3\frac{\frac{z_a}{z_b}+\frac{z_b}{z_a}}{\(\(\frac{z_a}{z_b}\)^2+\(\frac{z_b}{z_a}\)^2\)^2}=\frac{L^2}{(8\pi G_5\mu)^2(1-\lam_1\lam_2)}
\end{array}
\right.
\ee
where the first equation follows from $\frac{C_1C_2}{D_1D_2}=1$ and the
second equation can be obtained from $C_1C_2=\frac{L^2}{4}$.
The first equation can be used to give 
$\frac{z_a}{z_b}+\frac{z_b}{z_a}=\sqrt{\frac{(\lam_1+1)(\lam_2+1)}{\lam_1\lam_2}}$. 
Combining this with the second equation, we can
 express $\mu$ as a function of $\lam_1$ and $\lam_2$, which in terms of
variable $F=\lam_1\lam_2$ and $r=\frac{(\lam_1+\lam_2)^2}{\lam_1\lam_2}=\frac{(z_1^4+z_2^4)^2}{z_1^4z_2^4}$ reads:

\be\label{rF}
\frac{(8\pi G_5\mu)^2(z_1z_2)^3}{L^6}=\frac{F^{3/4}}{1-F}\(\frac{\sqrt{rF}+1}{F}+1\)^{1/2}\(\frac{\sqrt{rF}+1}{F}-3\)
\ee
 Note that $r$ depends on the degree of the asymmetry of the collision, for $z_1=z_2$ one has $r=4$. Let us thus fix $r$ and study the RHS
 of the (\ref{rF}) as a function of the other variable $F$, to be called   $A(r,F)$. 
From the second equation of (\ref{lam}), we know $F<1$ and by definition $F>0$.
%Define the r.h.s. of (\ref{rF}) as $A(r,F)$. 
For a given $r$, we have the following limits: as $F\rightarrow0$,
$A\rightarrow F^{-\frac{3}{4}}$ and as $F\rightarrow1$, 
$A\rightarrow\frac{1}{1-F}(\sqrt{r}+2)^{\frac{1}{2}}(\sqrt{r}-2)$.  Unless $r=4$ (the symmetric case), in both limits the function  tends to
positive infinity. Therefore a minimum must exist at certain $F=F_{min}$, 
which gives rise to the critical condition we are looking for. Fig.\ref{ArF} contains a plot
of $A$ as a function of $F$ at several $r$.

\begin{figure}[t]
\includegraphics[width=0.5\textwidth]{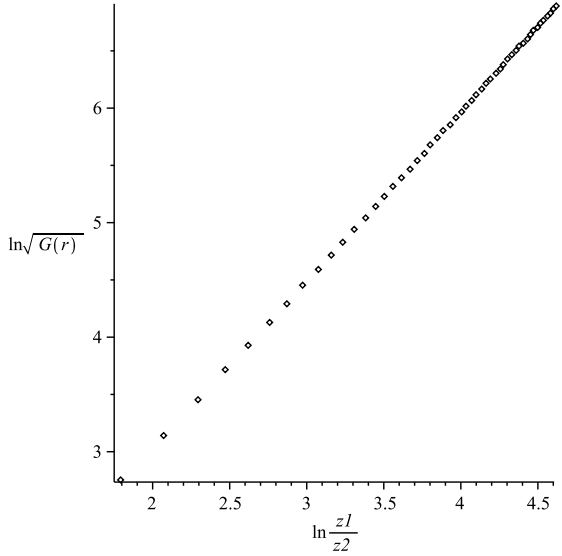}
\caption{\label{power}A log-log plot of $\sqrt{G(r)}$ versus $\frac{z_1}{z_2}$}
\end{figure}

The extremum of $A(r,F)$ is
found to be the roots of the following equation:
\be\label{extremum}
-3F-4rF^2+3\sqrt{rF}+3F^3+13F^2-6F\sqrt{rF}+3F^2\sqrt{rF}+3=0
\ee
It is not difficult to locate the minimum of $A(r,F)$ numerically, which
gives rise to a critical condition for the collision energy:
\be\label{critical_mu}
\frac{4\pi^2}{N_c^2}\mu(z_1z_2)^{3/2}\ge\sqrt{G(r)}
\ee

where we have used $G_5=\frac{\pi L^3}{2N_c^2}$. $G(r)$ is the minimum of
$A(r,F)$ at a given $r$. For $r=4(z_1=z_2)$, $A(r,F)$ has a minimum at $F\rightarrow1$: $G(r)=4$. We recover the critical condition for the symmetric collision
(\ref{critical_z0}). For general $r>4$,
we find $G(r)$ numerically and as $\frac{z_1}{z_2}$ grows, $\sqrt{G(r)}$
has a power like asymptotics $\sqrt{G(r)}\sim\(\frac{z_1}{z_2}\)^{3/2}$.
Fig.\ref{power} shows a the power law dependence of $G(r)$ on $\frac{z_1}{z_2}$.
The power measured by the slope in the log-log plot is approximately $1.5$.

The asymptotic power law behavior of $G(r)$ can be obtained analytically. 
We note the
root of (\ref{extremum}) corresponding to the minimum of $A(r,F)$ goes to
zero as $r\rightarrow\infty$. As the result, (\ref{extremum}) simplifies to
$-4rF^2+3\sqrt{rF}=0$, which is solved by 

\be
F=\(\frac{3}{4}\)^{2/3}r^{-\frac{1}{3}}+\cdots
\ee
where $\cdots$ denotes subleading terms. 
Substituting the root to $A(r,F)$, we obtain $G(r)=r^{\frac{3}{4}}+\cdots$. Combined
with the definition of $r$, we indeed have:

\be
\sqrt{G(r)}=\(\frac{z_1}{z_2}\)^{\frac{3}{2}}+\cdots
\ee
In the limit $r\gg1$($z_1\gg z_2$), the critical condition simplifies to
\be\label{asymp}
\frac{16\pi^2}{N_c^2}\mu z_2^3\ge 1
\ee

We would like to point out the non-uniqueness
of the trapped surface, as first remarked by Eardley and Giddings
\cite{eardley}, the
unusual boundary value problem defining the trapped surface could have
multiple solutions. We will see it is indeed the case in our wall-on-wall
collision.\footnote{Apart from this, there is also 
the foliation dependence of the trapped surface, which we do not discuss}.

Suppose we have the energy of the shock wave well above the critical value,i.e.
$A_0\equiv\frac{4\pi^2}{N_c^2}\mu(z_1z_2)^{3/2}\gg G(r)$. We know from the
previous analysis that

\be
&&A(r,F)\rightarrow F^{-3/4}\quad\text{as}\quad F\rightarrow 0 \no
&&A(r,F)\rightarrow \frac{1}{1-F}(\sqrt{r}+2)^{\frac{1}{2}}(\sqrt{r}-2)\quad
\text{as}\quad F\rightarrow 1 \nonumber
\ee

This allows two solutions $F=A_0^{-4/3}+\cdots$ and 
$F=1-\frac{(\sqrt{r}+2)^{\frac{1}{2}}(\sqrt{r}-2)}{A_0}+\cdots$. 
Without explicit solution of the trapped surface, 
we can compare the area of two corresponding trapped surface, which is
related to the entropy production per transverse area\cite{LS}:

\be
&&s=\frac{N_c^2}{2\pi}\(\frac{1}{z_a^2}-\frac{1}{z_b^2}\) \no
&&\,=\(2\pi N_c^2\mu\)^{1/3}A_0^{-1/3}F^{-1/4}\(1+\sqrt{rF}-3F\)^{1/2}
\ee

With the former solution, we have $s=\(2\pi N_c^2\mu\)^{1/3}+\cdots$, while 
the latter solution gives rise to 
$s=\(2\pi N_c^2\mu\)^{1/3}A_0^{-1/3}(\sqrt{r}-2)^{1/2}+\cdots$. In the limit
$A_0\rightarrow\infty$, the former trapped surface has a much greater area
than the latter. Therefore we choose the former as
the ``outermost'' trapped surface. This branch of solution is precisely the one
used in \cite{AAtherm} for a comparison of sourced shock wave and source-free
shock wave.

\begin{figure}[t]
\includegraphics[width=0.8\textwidth]{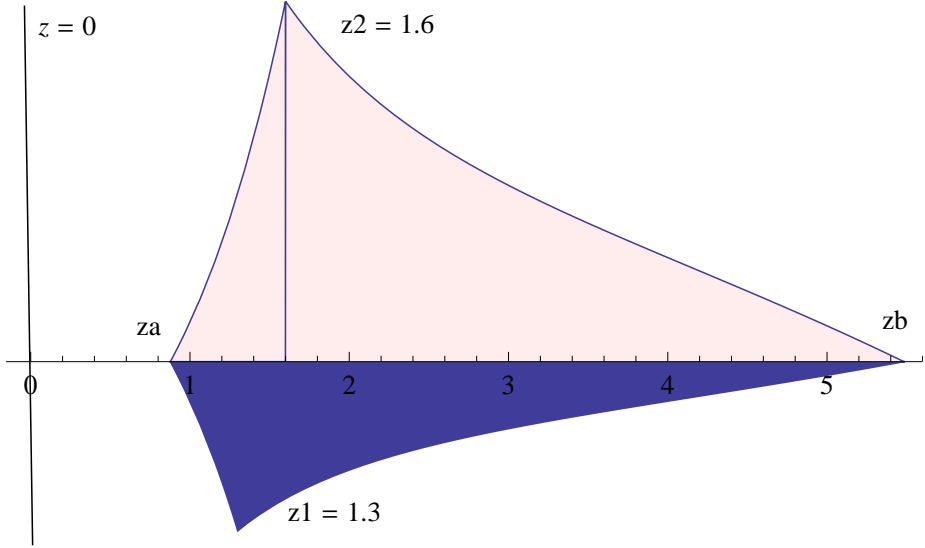}
\caption{\label{psi}(color online)A view of the outer-most trapped surface 
formation in a wall-on-wall collision with two sources at different depths.
The pink and blue area indicate the growth of the trapped surface $\Psi$ 
in the bulk. The sources of the shock waves lie at $\frac{z_1}{L}=1.3$ and 
$\frac{z_2}{L}=1.6$, and the energy density is fixed by
$\frac{(8\pi G_5\mu)^2(z_1z_2)^3}{L^3}=20$.
The trapped surface at the collision point is bounded by $z_a$ and $z_b$}
\end{figure}

A view of the trapped surface formation is included in Fig.\ref{psi}.
We note the trapped surface only starts to appear when the wave fronts of 
the shocks are separated by finite distance, unlike the situation
in point shock wave collision, where the trapped surface starts to grow
even for infinitely separated shock waves.
Furthermore, the origin of the critical condition in wall-on-wall collision
is different from that of point shock wave collision :
The latter can be
understood as a constraint on angular momentum for the shock wave pair 
to form a AdS-Kerr black hole. The energy dependence of the critical impact
parameter obeys an asymptotic
 power law, with the power extracted numerically in \cite{LS}
to be $0.37$, and argued by Gubser, Pufu and Yarom to be $1/3$\cite{GPY2}. 
The power $1/3$ was confirmed later in a detailed numerical analysis\cite{mozo}.
%Indeed, the
%estimate from angular momentum and center of mass gives the power $1/3$\cite{}.
The critical condition for wall-on-wall collision (\ref{critical_mu}) 
has no analogy here, as the
shock wave pair does not have an obvious angular momentum. 
In the next section, we will compute the NLO stress tensor in the dual field
theory, which will help us to understand the origin of the critical condition.

\section{The NLO stress tensor after the collision}

The Einstein equation
in the presence of the cosmological constant is given by:

\be
G_{\mu\nu}+6g_{\mu\nu}=-8\pi G_5J_{\mu\nu}
\ee

with $J_{\mu\nu}$ the 5-dimensional source in the bulk. We choose to work
with an alternative form of the Einstein equation:

\be\label{EE}
R_{\mu\nu}-4g_{\mu\nu}=-8\pi G_5S_{\mu\nu}
\ee
where $S_{\mu\nu}=J_{\mu\nu}-\frac{1}{3}Jg_{\mu\nu}$. We have set the AdS radius
$L=1$. Since the relevant scales are completely fixed by $\mu$ and $z_i$,
we expect $L$ will be absent in the final result of the stress tensor.
The equation (\ref{EE}) to the first order in the amplitude of the shock
wave is:

\be\label{LO_EE}
R_{\mu\nu}^{(1)}-4g_{\mu\nu}^{(1)}=-8\pi G_5S_{\mu\nu}^{(1)}=-8\pi G_5(J_{\mu\nu}^{(1)}-\frac{1}{3}J^{(1)}g_{\mu\nu}^{(0)})
\ee
where the upper index denotes the order of the quantity with respect to the
amplitude of the shock wave. e.g. $g_{\mu\nu}^{(0)}$ is the pure AdS metric.
Before the collision the superposition of two shock waves are solution to this 
equation. Their contribution to the source is of first order:

\be
&&8\pi G_5J_{++}^{(1)}=-\frac{1}{2}\nabla^2\phi_1 \no
&&8\pi G_5J_{--}^{(1)}=-\frac{1}{2}\nabla^2\phi_2 
\ee

After the shock waves pass through each other, the $source$ of either shock wave
feels the $field$ of the other shock wave and deviates from its original 
trajectory. The deviation gives rise to the second order correction to 
the source: $J_{\mu\nu}^{(2)}$. On the other hand, since the superposition 
of two shock waves
does not satisfy Einstein equation after the collision, thus a nonvanishing
Ricci tensor is expected from the superposition. 
We collect this contribution into $R_{\mu\nu}^{(1,1)}$, which
 can be interpreted as the interaction of the shock wave $fields$. Combing
the two contributions, the second order Einstein equation now takes the
following form:

\be\label{NLO_EE}
R_{\mu\nu}^{(2)}+R_{\mu\nu}^{(1,1)}-4g_{\mu\nu}^{(2)}=-8\pi G_5S_{\mu\nu}^{(2)}=
-8\pi G_5(J_{\mu\nu}^{(2)}-J^{(2)}g_{\mu\nu}^{(0)}-J^{(1)}g_{\mu\nu}^{(1)})
\ee

Our contracted source of order $k$ is always defined as 
$J^{(k)}=J_{\mu\nu}^{(k)}g^{\mu\nu}{}^{(0)}$.

The calculation of 
$J_{\mu\nu}^{(2)}$ needs some explanations. Since our wall shock wave has
trivial dependence on transverse coordinates, the problem gets simplified a lot.
The sources can only move in $z$, and  the determination of
their trajectory after the collision is subtle. In general it depends on 
the equation of state of the extended source itself. We will assume the action
due to the shock wave source is of Nambu-Goto type, which is proportional
to the invariant area of the extended source.

%We can choose the worldvolume coordinates of the source of the shock wave as
%$\lam,x_1,x_2$, with $\lam$ to be specified later. The trajectory of the
%wall source can be specified by $\xp=X^+(\lam),\,\xm=X^-(\lam),\,z=Z(\lam)$.

%Suppose the covariant source $J^{\mu\nu}$ due to one the shock wave 
%sources is parametrized by

%\be
%J^{\mu\nu(2)}=\#\int d\lam u^{\mu}u^{\nu}\dlt(\xp-X^+(\lam))\dlt(\xm-X^-(\lam))\dlt(z-Z(\lam))
%\ee
%with $u^{\mu}=\frac{dx^{\mu}}{d\lam}\quad\mu=+,-,z$. $\#$ can be some function
%of $\xp,\,\xm,\,z$.

Calculating the geodesic of one shock wave $source$ in the background
of the other shock wave $field$(details can be found in the appendix). The 
second order source is given by:

\be\label{J2}
&&8\pi G_5J_{++}^{(2)}=\frac{1}{2}(\int\phi_2 d\xm\del_+\nabla^2\phi_1+\frac{1}{2}\int d\xm\int d\xm\del_z\phi_2 \del_z\nabla^2\phi_1) \no
&&8\pi G_5J_{--}^{(2)}=\frac{1}{2}(\int\phi_1 d\xp\del_-\nabla^2\phi_2+\frac{1}{2}\int d\xp\int d\xp\del_z\phi_1 \del_z\nabla^2\phi_2) \no
&&8\pi G_5J_{+-}^{(2)}=\frac{1}{2}(\phi_1\nabla^2\phi_2+\phi_2\nabla^2\phi_1) \no
&&8\pi G_5J_{+z}^{(2)}=\frac{1}{2}\int\del_z\phi_2d\xm\nabla^2\phi_1 \no
&&8\pi G_5J_{-z}^{(2)}=\frac{1}{2}\int\del_z\phi_1d\xm\nabla^2\phi_2 
\ee

We can check the following relations to the second order:

\be\label{conservation}
&&(\nabla^{\mu}J_{\mu\nu})^{(2)}=\nabla^{\mu(0)}J_{\mu\nu}^{(2)}+\nabla^{\mu(1)}J_{\mu\nu}^{(1)} \nonumber\\
&&=-2z^4\dlt(z-z_1)\dlt(z-z_2)(\theta(\xp)\dlt(\xm)+\theta(\xm)\dlt(\xp))=0 \\
&&(g^{\mu\nu}J_{\mu\nu})^{(2)}=g^{\mu\nu(0)}J_{\mu\nu}^{(2)}+g^{\mu\nu(1)}J_{\mu\nu}^{(1)}=0
\ee

The first one is the conservation of the source, which is a necessary condition
for the consistency of Einstein equation. The second traceless condition
allows us to simplify the RHS of (\ref{NLO_EE}). Moving
The Ricci tensor quadratic in the first order metric $R_{\mu\nu}^{(1,1)}$ to the
RHS and noting the tracelessness of $J^{(2)}=J^{(1)}=0$, 
we obtain the reshuffled Einstein equation for the second order corrections only
\be\label{EE_reshuffle}
R_{\mu\nu}^{(2)}-4g_{\mu\nu}^{(2)}=-8\pi G_5J_{\mu\nu}^{(2)}-R_{\mu\nu}^{(1,1)}=-8\pi G_5\bar{J}_{\mu\nu}^{(2)}
\ee
where we have defined the effective source 
$\bar{J}_{\mu\nu}^{(2)}=J_{\mu\nu}^{(2)}+\frac{1}{8\pi G_5}R_{\mu\nu}^{(1,1)}$

It is easy to work out $R_{\mu\nu}^{(1,1)}$ for the case of wall shock waves
and we obtain the effective source as:

\be\label{barJ2}
&&8\pi G_5\bar{J}_{++}^{(2)}=\frac{1}{2}\int d\xm\phi_2\del_+\nabla^2\phi_1-\frac{1}{4}\int d\xm\int d\xm\del_z\phi_2\del_z\nabla^2\phi_1 \no
&&8\pi G_5\bar{J}_{--}^{(2)}=\frac{1}{2}\int d\xp\phi_1\del_-\nabla^2\phi_2-\frac{1}{4}\int d\xp\int d\xp\del_z\phi_1\del_z\nabla^2\phi_2 \no
&&8\pi G_5\bar{J}_{+-}^{(2)}=\frac{1}{2}(\phi_1\nabla^2\phi_2+\phi_2\nabla^2\phi_1)-\del_+\phi_1\del_-\phi_2+\del_z\phi_1\del_z\phi_2-\frac{1}{z}(\phi_1\del_z\phi_2+\phi_2\del_z\phi_1) \no
&&8\pi G_5\bar{J}_{+z}^{(2)}=\frac{1}{2}\int d\xm\del_z\phi_2\nabla^2\phi_1-\del_+\phi_1\del_z\phi_2 \no
&&8\pi G_5\bar{J}_{-z}^{(2)}=\frac{1}{2}\int d\xp\del_z\phi_1\nabla^2\phi_2-\del_-\phi_2\del_z\phi_1 \no
&&8\pi G_5\bar{J}_{\perp\perp}^{(2)}=\frac{2}{z}(\phi_1\del_z\phi_2+\phi_2\del_z\phi_1) \no
&&8\pi G_5\bar{J}_{zz}^{(2)}=-2(\phi_2\del_z^2\phi_1+\phi_1\del_z^2\phi_2)+\frac{2}{z}(\phi_2\del_z\phi_1+\phi_1\del_z\phi_2)-2\del_z\phi_1\del_z\phi_2 
\ee

From here on, we can use the method developed in \cite{dipole1,dipole2} to
compute the stress tensor on the boundary field theory to the NLO. The procedure
is to first obtain the reshuffled source $s_{mn}$ defined as
\be
s_{mn}^{(2)}=\bar{J}_{mn}^{(2)}-\int_0^z\(\bar{J}_{zm,n}^{(2)}+\bar{J}_{zn,m}^{(2)}\)dz+\frac{1}{2}h_{,m,n}+\frac{1}{2z}\eta_{mn} h_{,z}
\ee
with 
\be
h=\frac{1}{3}\int_0^z dz\cdot z
\(\bar{J}_{zz}^{(2)}-\eta^{mn}\bar{J}_{mn}^{(2)}+2\int_0^z dz \(-\eta^{mn}\bar{J}_{zm,n}^{(2)}\)\)
\ee

In our particular case, $h$ is given by:
\be\label{h}
8\pi G_5h=-4\phi_1\phi_2+4\int dz\cdot z\int dz\frac{1}{z}\del_z\phi_1\del_z\phi_2
\ee

The reshuffled source takes the following form:
\be\label{s2}
&&8\pi G_5s_{++}=\frac{1}{2}\int d\xm\phi_2\del_+\nabla^2\phi_1+\frac{1}{4}\int d\xm\int d\xm\del_z\phi_2\del_z\nabla^2\phi_1 \no
&&-\int d\xm\int dz\del_z\phi_2\del_+\nabla^2\phi_1-2\int dz\phi_2\del_z\del_+^2\phi_1+2\int dz\cdot z\int dz\frac{1}{z}\del_+^2\del_z\phi_1\del_z\phi_2 \no
&&8\pi G_5s_{--}=\frac{1}{2}\int d\xp\phi_1\del_-\nabla^2\phi_2+\frac{1}{4}\int d\xp\int d\xp\del_z\phi_1\del_z\nabla^2\phi_2 \no
&&-\int d\xp\int dz\del_z\phi_1\del_-\nabla^2\phi_2-2\int dz\phi_1\del_z\del_-^2\phi_2+2\int dz\cdot z\int dz\frac{1}{z}\del_-^2\del_z\phi_2\del_z\phi_1 \no
&&8\pi G_5s_{+-}=\del_z\phi_1\del_z\phi_2-2\del_+\phi_1\del_-\phi_2+\frac{1}{2}\int dz(\phi_1\del_z\nabla^2\phi_2+\phi_2\del_z\nabla^2\phi_1) \no
&&+2\int dz\cdot z\int dz\frac{1}{z}\del_+\del_z\phi_1\del_-\del_z\phi_2-\int dz\frac{1}{z}\del_z\phi_1\del_z\phi_2 \no
&&8\pi G_5s_{\perp\perp}=2\int dz\frac{1}{z}\del_z\phi_1\del_z\phi_2
\ee

Separating the derivatives of $\xp,\,\xm$ and derivative of $z$,
we obtain:
\be\label{s2_bar}
&&8\pi G_5s_{++}= \frac{1}{2}\bph_2\nabla^2\bph_1\theta(\xm)\dlt'(\xp)+\frac{1}{4}\bph_2'\nabla^2\bph_1'\xm\theta(\xm)\dlt(\xp)-2\int dz\bph_1'\bph_2\dlt''(\xp)\dlt(\xm) \no
&&+2\int dz\cdot z\int dz\frac{1}{z}\bph_1'\bph_2'\dlt''(\xp)\dlt(\xm)-\int dz\bph_2'\nabla^2\bph_1\theta(\xm)\dlt'(\xp)\no
&&8\pi G_5s_{--}= \frac{1}{2}\bph_1\nabla^2\bph_2\theta(\xp)\dlt'(\xm)+\frac{1}{4}\bph_1'\nabla^2\bph_2'\xp\theta(\xp)\dlt(\xm)-2\int dz\bph_2'\bph_1\dlt''(\xm)\dlt(\xp) \no
&&+2\int dz\cdot z\int dz\frac{1}{z}\bph_1'\bph_2'\dlt''(\xm)\dlt(\xp)-\int dz\bph_1'\nabla^2\bph_2\theta(\xp)\dlt'(\xm) \no
&&8\pi G_5s_{+-}= \frac{1}{2}\int dz(\bph_1\nabla^2\bph_2'+\bph_2\nabla^2\bph_1')\dlt(\xp)\dlt(\xm)+\bph_1'\bph_2'\dlt(\xp)\dlt(\xm) \no
&&-2\bph_1\bph_2\dlt'(\xp)\dlt'(\xm)+2\int dz\cdot z\int\frac{1}{z}\bph_1'\bph_2'\dlt'(\xp)\dlt'(\xm)-\int dz\frac{1}{z}\bph_1'\bph_2'\dlt(\xp)\dlt(\xm)\no
&&8\pi G_5s_{\perp\perp}= 2\int dz\frac{1}{z}\bph_1'\bph_2'\dlt(\xp)\dlt(\xm)
\ee

where $\phi_1(\xp,z)=\bar{\phi_1}(z)\dlt(\xp)$ and 
$\phi_2(\xm,z)=\bar{\phi_2}(z)\dlt(\xm)$. In (\ref{s2_bar}), all primes
are ordinary derivatives. The explicit forms of $\bph_1$ and $\bph_2$ are
given by:

\be\label{bph}
&&\bph_1=4\pi G_5\mu_1\frac{z_1^4}{L^3}\frac{z^4-(z^4-z_1^4)\theta(z-z_1)}{z_1^4} \\
&&\bph_2=4\pi G_5\mu_2\frac{z_2^4}{L^3}\frac{z^4-(z^4-z_2^4)\theta(z-z_2)}{z_2^4}
\ee

The reshuffled source (\ref{s2_bar}) will be convoluted with a bulk to
boundary propagator in AdS. 
Such propagator has been built in various applications
of AdS/CFT, e.g.\cite{giddings,AKT,dipole2,GY,sync}. Propagators in an AdS shock
wave background were found in \cite{costa,brower}.
We will use a slightly different propagator from the above. The propagator
takes the following form in the lightcone coordinates:

\be\label{prop}
P_R=\frac{\theta(\xp-\xp{}'+\xm-\xm{}')}{2\pi}\[\frac{\dlt'''(z-w)^{1/2})}{8w^3}+\frac{3\dlt''(z-w)}{8w^4}+\frac{3\dlt'(z-w)}{8w^5}\]
\ee
with $w=\sqrt{(\xp-\xp{}')(\xm-\xm{}')-(\vx_\perp-\vx_\perp')^2}$.

The details of the propagator are included in the appendix. Since we are 
dealing with wall sources, which do not depend on $x_\perp'$, we can perform
the integral with respect to the transverse coordinate $x_\perp'$. By repeated 
use of integration by parts, we end up with a concise form:

\be\label{prop_int}
&&\int d^2x_\perp'P_R=2\pi\int_0^\infty dx_\perp'\cdot x_\perp'P_R= 
\theta(\xp-\xp{}'+\xm-\xm{}')\times \no
&&\[\frac{\dlt'(z-\sqrt{(\xp-\xp{}')(\xm-\xm{}')})}{8{(\xp-\xp{}')(\xm-\xm{}')}^{3/2}}+\frac{\dlt''(z-\sqrt{(\xp-\xp{}')(\xm-\xm{}')})}{8(\xp-\xp{}')(\xm-\xm{}')}\]
\ee

The final task is to convolute the source (\ref{s2_bar}) with the integrated
propagator (\ref{prop_int}). Due to the presence of the delta function, 
the integration in $z$ is trivial.
We are only left with integration of $\xp{}'$ and $\xm{}'$. Completing the
integrals, we obtain as the final results:

\be\label{Tnlo}
&&T_{++}^{NLO}=\frac{8\pi^2 G_5\mu_1\mu_2}{N_c^2}\[-\xm{}^2\theta(z_2-\tau)\theta(\tau)+\frac{\xm z_2^3}{2\xp}\dlt(z_2-\tau)\] \no
&&T_{--}^{NLO}=\frac{8\pi^2 G_5\mu_1\mu_2}{N_c^2}\[-\xp{}^2\theta(z_2-\tau)\theta(\tau)+\frac{\xp z_2^3}{2\xm}\dlt(z_2-\tau)\] \no
&&T_{+-}^{NLO}=\frac{1}{2}T_{\perp\perp}^{NLO}=\frac{8\pi^2 G_5\mu_1\mu_2}{N_c^2}\[2\tau^2\theta(z_2-\tau)\theta(\tau)-\frac{z_2^3}{2}\dlt(z_2-\tau)\]
\ee
where $\tau=\sqrt{\xp\xm}$ is the proper time. This is the main result of this chapter.

(Few technical comments on the derivation: We have also used
the distributional relations in the final results 
$\dlt^{(n)}(x)f(x)=(-1)^nf^{(n)}(x)\dlt(x)$. In doing this, we have treated
$\tau$ and $\frac{\xp}{\xm}$ as separate variables.
It is however necessary to keep in mind one subtlety.
We have assumed the source has a series expansion near the boundary $z=0$,
in the derivation of the propagator. 
As our source contains delta functions and Heaviside 
theta functions, (\ref{Tnlo}) is obtained with a particular representation
of them and the limit is taken in the final results.)

Several more general comments on the result, the NLO stress tensor (\ref{Tnlo}), are in order:

i) The NLO stress tensor is conserved and traceless 
$\del^mT_{mn}^{NLO}=0,\,\eta^{mn}T_{mn}^{NLO}=0$. The
presence of the delta function is necessary for the conservation relation.
As conjectured in \cite{AAtherm}, 
The limit $z_2\rightarrow\infty$ of our results should recover 
the NLO stress tensor in the collision of sourceless shock wave\cite{romatschke,AKT}. We can see it is indeed the case as $\theta(z_2-\tau)=1,\,\dlt(z_2-\tau)=0$.

ii) (\ref{Tnlo}) is actually boost invariant. In a comoving frame with
coordinate $\tau=\sqrt{\xp\xm}$ and $\eta=\frac{1}{2}\ln\frac{\xm}{\xp}$, the
NLO stress tensor takes the following form:

\be
&&T_{\tau\tau}^{NLO}=\frac{8\pi^2 G_5\mu_1\mu_2}{N_c^2}\[2\tau^2\theta(z_2-\tau)\theta(\tau)\] \no
&&T_{\eta\eta}^{NLO}=\frac{8\pi^2 G_5\mu_1\mu_2}{N_c^2}\[-6\tau^4\theta(z_2-\tau)\theta(\tau)+2\tau^2 z_2^3\dlt(z_2-\tau)\] \no
&&T_{\perp\perp}^{NLO}=\frac{8\pi^2 G_5\mu_1\mu_2}{N_c^2}\[4\tau^2\theta(z_2-\tau)\theta(\tau)-z_2^3\dlt(z_2-\tau)\]
\ee

The boost invariance is a special property of the NLO stress tensor, 
which is symmetric under the exchange of the two shock waves. Since
we are colliding asymmetric nucleus, we expect higher order correction
should violate boost invariance.
%This is important result, confirming that the initial conditions for the hydrodynamical evolution should be of the
%so called Bjorken (rather than Landau) type.

iii) It is interesting to note that the NLO stress tensor does not depend on
$z_1$. It suggests the NLO stress tensor for
 collision of two nucleus with different saturation scales does not feel 
the softer saturation scale $\frac{1}{z_1}$.

iv) The appearance of the Heaviside theta
function is of particular interest. It encodes information on thermalization. 
As the LO
stress tensor $T_{++}^{LO}=\mu_1\dlt(\xp),\,T_{--}^{LO}=\mu_2\dlt(\xm)$ has
a simple interpretation as nucleus moving on the lightcone. The NLO
stress tensor (\ref{Tnlo}) tells us matter created in the collision
is only nonvanishing when $0<\tau<z_2$. At time $t>z_2$, 
matter created in the
collision separates into two pieces $z_2<x_3<t$ and $-t<x_3<-z_2$. Presumably
higher order correction is needed to fill the gap. This also suggest the NLO
result is insufficient to provide an initial condition for hydrodynamics.

v) Comparing the normalization
of the delta functions in the LO and NLO stress tensor, we conclude
the perturbation should break down when $\mu\lesssim 8\pi G_5\mu^2z_2^3$,
which is precisely the critical condition (\ref{asymp}).

Therefore the field theory interpretation of the thermalization
condition is understood as the breaking down of perturbative treatment. Presumably
the combined effect of all the gravitons should be included in further evolution
of the trapped surface, from its position at time zero discussed at the beginning of the paper.
%a resummation of graviton exchanges is needed to reach thermalization.
%In order for the equilibrarion
%to happen, we would have to require the perturbation in the amplitude
%of the shock waves breaks down at $t\lesssim z_2$. The breaking down
%of the perturbation should happen when the sources of the shock wave
%deviate from their original paths significantly, which gives rise to the
%following condition for matter equilibration:

%\be\label{deviation}
%t\sim\frac{\# z_2}{u^z}=\frac{\#z_2}{32\pi G_5\mu_1 z_2^3}\lesssim z_2
%\ee

Alternatively, we can take a bulk point of view: The perturbation breaks
down when the sources of the shock wave, originally moving at constant
radial position, deviate significantly in the radial direction.
Specially, we have worked this out in the Appendix A. With $\mu_1=\mu_2=\mu$,
the sources of shock wave gain velocities

\be
&&u^z_1=\int d\xm\frac{\del_z\phi_2}{2}\vert_{z=z_1}=0 \no
&&u^z_2=\int d\xp\frac{\del_z\phi_1}{2}\vert_{z=z_2}=\frac{8\pi^2\mu}{N_c^2}z_2^3\theta(\xp)
\ee
after the collision. Due to the special profile of the shock waves, the 
source deeper in the bulk does not shift its path in the NLO computation.
The perturbation breaks down when $u^z_2\lesssim 1$, which again is consistent
with the critical condition (\ref{asymp}) when $\frac{z_1}{z_2}\gg 1$.

The improved understanding of the critical condition leads to the
following prediction: In the collision of two nucleus with the same energy 
density but different saturation scale, the thermalization condition is
insensitive to the softer saturation scale. The energy density has to exceed
certain critical value set by the harder saturation scale as (\ref{asymp})
in order to reach thermalization.
%We have set $\mu_1=\mu_2=\mu$ for
%comparison with (\ref{critical_mu}). (\ref{deviation}) can be rewritten as:

%\be\label{asymp}
%\frac{16\pi^2}{N_c^2}\mu(z_1z_2)^{3/2}\gtrsim\#\(\frac{z_1}{z_2}\)^{3/2}
%\ee

%While the power of $(z_1z_2)$ is fixed on dimensional ground. While the power
%of $\frac{z_1}{z_2}$ is a nontrivial prediction from the above critierium.
%It indeed agrees with the result of the trapped surface analysis 
%(\ref{critical_mu})
%when $\frac{z_1}{z_2}\gg 1$. Therefore, we conclude the critical condition
%arises from the breaking down of perturbation in a time scaled set by
%the harder saturation scale.

%\be\label{delta_reg}
%&&\theta(x)=\frac{1}{2\pi}(i\ln(x+i\eps)-i\ln(x-i\eps)) \no
%&&\dlt(x)=\frac{1}{2\pi}(\frac{i}{x+i\eps}-\frac{i}{x-i\eps}) \no
%&&\dlt'(x)=\frac{1}{2\pi}(-\frac{i}{(x+i\eps)^2}+\frac{i}{(x-i\eps)^2})
%\ee

%Working with the regularized form of the source, we obtain in the final
%results also forms of (\ref{delta_reg})

\section{Discussion}

In this work, we have constructed the trapped surface in a wall-on-wall
collision, which is used to model collisions of nucleus with different 
saturation scales. We have derived a critical condition for 
matter equilibration in nucleus collisions.
The condition
(\ref{critical_mu}) is set by the saturation scales of both nucleus.
The critical energy scales as the ratio of the saturation scales approximately
by a power law, with the power $3/2$. The approximate power law indicates
the critical energy is insensitive to the softer saturation scale.
We have also observed a non-uniqueness of the trapped surface when the
energy density is beyond the critical value.
The outer-most trapped surface is selected for an estimate of the
entropy production.

We have computed the NLO stress tensor on the boundary. The result turns out
to be independent on the soft saturation scale $1/z_1$. Based on the NLO
results, we propose the critical condition corresponds to the
breaking down of the perturbation, i.e. when the LO and NLO correction become
comparable. The criterion reproduces the critical condition (\ref{asymp}).
On the other hand, the critical condition is also understood in terms of
bulk physics. The breaking down of the perturbation is encoded in the
condition when the sources of the shock wave gain significant deviation
in its velocity after the collision. This also leads to the correct
critical condition (\ref{asymp}). 
While in the NLO, no dependence on $z_1$ is observed, it must show up beyond
NLO, as the source deeper in the bulk will also deviate from
its original path, giving rise to correction to (\ref{asymp}). 
It is tempting to see how this shows up in higher order computation.

Finally we stress the physics of critical condition for matter equilibration 
is very
different from the counterpart in the collision of point source shock wave. 
In terms of the gravity dual, the latter originates from the constraint on
the angular momentum possessed by a pair of black holes in order for the
merging to be possible. The critical condition for wall shock wave collision
can be understood as the breaking down of the perturbative calculation.
In the dual field theory, it manifests as a constraint on the 
collision energy for given parton saturation scale.

%%%%%%%%%%%%%%% ack %%%%%%%%%%%%%%%%%
\noindent{\large \bf Acknowledgments} \vskip .35cm We thank
G. Beuf and A. Taliotis for helpful discussions. S.L. thank
Erwin Sch\"odinger International Institute and 
KITP China for hospitalities during the initial and final stages of the paper.
The work of S.L. was
supported by Alexander von Humboldt Foundation. The work of E.S.
was partially
supported by the US-DOE grants DE-FG02-88ER40388 and
DE-FG03-97ER4014.

\appendix
\section{NLO source in shock wave collision}

\subsection{Point source shock wave in $AdS_3$}
It is helpful to look at collision of point shock wave in $AdS_3$ first. The LO
metric is given by:

\be
ds^2=-\frac{-d\xp d\xm+dz^2+\phi_1 d\xp{}^2+\phi_2 d\xm{}^2}{z^2}
\ee

The shock wave profiles $\phi_1$ and $\phi_2$ are normalized as:

\be
&&\nabla^2\phi_1=-16\pi G_5\dlt(\xp)\dlt(z-z_1) \\
&&\nabla^2\phi_2=-16\pi G_5\dlt(\xm)\dlt(z-z_2)
\ee

where $\nabla^2=\del_z^2-\frac{1}{z}\del_z$ is a Laplacian operator. The NLO
source arises from the deviation of the path of one shock wave source in
the presence of the other. For point source, the null geodesic equation
 is given by:

\be
\frac{du^\mu}{d\lam}+\Gamma^\mu_{\alpha\beta}u^{\alpha}u^{\beta}=0
\ee

For source of shock wave 1 before the collision, $\xm$ can be chosen as 
the affine parameter $\lam$,
thus $u^-=1,\,u^+=u^z=0$. After the collision, the geodesic to the first order
in the shock wave amplitude is as follows:

\be
&&\frac{du^+}{d\lam}=\del_-\phi_2 \label{up}\\
&&\frac{du^z}{d\lam}=\frac{1}{z}u^++\frac{z^2}{2}\del_z(\frac{\phi_2}{z^2}) \label{uz}\\
&&\frac{du^-}{d\lam}=\frac{2}{z}u^z \label{um}
\ee

Assuming $\lam=\xm$ holds after the collision, we find from (\ref{up}) and 
(\ref{uz}) that $u^+=\phi_2$ and $u^z=\int d\xm\frac{\del_z\phi_2}{2}$. However
we see it contradicts (\ref{um}) as $\frac{du^-}{d\xm}=0$. This indicates
that $\xm$ is no longer a good affine parameter, but to the order we are
interested, $u^+=\phi_2$ and $u^z=\int \frac{\del_z\phi_2}{2}d\xm$ remains
valid, as correction will be of higher order. Integrating once, we further
obtain $\xp=\int d\xm\phi_2$ and $z=z_1+\int d\xm\int d\xm\frac{\del_z\phi_2}{2}$.

The covariant source due to shock wave 1 has the general form:

\be
J^{\mu\nu}=\#u^\mu u^\nu\dlt(\xp-X^+(\xm))\dlt(z-Z(\xm))
\ee

where $X^+(\xm)$ and $Z(\xm)$ specifies the trajectory of the point source.
$\#$ can be some function of $\xp,\,\xm$ and $z$. 
Writing the LO covariant stress tensor is simply:

\be
&&8\pi G_5J^{--(1)}=-2z^4\nabla^2\phi_1 \\
&&8\pi G_5J^{++(1)}=-2z^2\nabla^2\phi_2
\ee

The NLO source comes from the correction to $u^\mu$ and $x^\mu$. Adding
the contributions from two shock waves, we obtain:

\be\label{ads3}
&&8\pi G_5J^{--(2)}=2z^4(\int d\xm\phi_2\del_+\nabla^2\phi_1+\frac{1}{2}\int d\xm\int d\xm\del_z\phi_2\del_z\nabla^2\phi_1) \no
&&8\pi G_5J^{++(2)}=2z^4(\int d\xp\phi_1\del_-\nabla^2\phi_2+\frac{1}{2}\int d\xp\int d\xp\del_z\phi_1\del_z\nabla^2\phi_2) \no
&&8\pi G_5J^{+-(2)}=-2z^4(\phi_1\nabla^2\phi_2+\phi_2\nabla^2\phi_1) \no
&&8\pi G_5J^{-z(2)}=-2z^4\cdot\frac{1}{2}\int d\xm\del_z\phi_2\nabla^2\phi_1 \no
&&8\pi G_5J^{+z(2)}=-2z^4\cdot\frac{1}{2}\int d\xp\del_z\phi_1\nabla^2\phi_2
\ee

The conservation of the source to the second order can be checked
$(\nabla_\mu T^{\mu\nu})^{(2)}=\del_\mu T^{\mu\nu(2)}+\Gamma^{\mu(0)}_{\mu\lam}T^{\lam\nu(2)}+\Gamma^{\nu(0)}_{\mu\lam}T^{\mu\lam(2)}+\Gamma^{\mu(1)}_{\mu\lam}T^{\lam\nu(1)}+\Gamma^{\nu(1)}_{\mu\lam}T^{\mu\lam(1)}=0$.

\subsection{Wall source shock wave in $AdS_5$}

Now we look at wall shock wave in $AdS_5$. The LO metric is given by:

\be
ds^2=\frac{-d\xp d\xm+dx_\perp^2+dz^2+\phi_1d\xp{}^2+\phi_2d\xm{}^2}{z^2}
\ee

The shock wave profiles $\phi_1$ and $\phi_2$ are normalized as
\be
&&\nabla^2\phi_1=-16\pi G_5\dlt(\xp)\dlt(z-z_1) \\
&&\nabla^2\phi_2=-16\pi G_5\dlt(\xm)\dlt(z-z_2)
\ee

The Laplacian operator becomes $\nabla=\del_z^2-\frac{3}{z}\del_z$ due to
the additional transverse directions. Being
different from the point source, the trajectory of the source is specified
by $X^\mu(\sigma)$ with $\sigma$ the worldvolume parameters. The induced
metric is given by:

\be
h_{\alpha\beta}=\frac{\del x^\mu}{\del \sigma^\alpha}\frac{\del x^\nu}{\del \sigma^\beta}g_{\mu\nu}=
\begin{pmatrix}
\frac{-u^+u^-+u^z{}^2}{z^2}& & \\
& \frac{1}{z^2}& \\
& & \frac{1}{z^2}
\end{pmatrix}
\ee

with $u^{\pm}=\frac{dx^\pm}{d\lam}$ and $u^z=\frac{dz}{d\lam}$. Assuming the
action of the shock wave depends on $\det h$ only, then the trajectory
can be effectively determined by considering a point source in the metric

\be
ds^2=\frac{-d\xp d\xm+dz^2+\phi_1 d\xp{}^2+\phi_2 d\xm{}^2}{z^6}
\ee

Working out the geodesic deviation, we find surprisingly that the trajectory
of the wall source is the same as point source in $AdS_3$. As a result,
the LO and NLO source are given by:

\be\label{ads5}
&&8\pi G_5J^{--(1)}=-2z^4\nabla^2\phi_1 \\
&&8\pi G_5J^{++(1)}=-2z^2\nabla^2\phi_2 \\
&&8\pi G_5J^{--(2)}=2z^4(\int d\xm\phi_2\del_+\nabla^2\phi_1+\frac{1}{2}\int d\xm\int d\xm\del_z\phi_2\del_z\nabla^2\phi_1) \\
&&8\pi G_5J^{++(2)}=2z^4(\int d\xp\phi_1\del_-\nabla^2\phi_2+\frac{1}{2}\int d\xp\int d\xp\del_z\phi_1\del_z\nabla^2\phi_2) \\
&&8\pi G_5J^{+-(2)}=-2z^4(\phi_1\nabla^2\phi_2+\phi_2\nabla^2\phi_1) \\
&&8\pi G_5J^{-z(2)}=-2z^4\cdot\frac{1}{2}\int d\xm\del_z\phi_2\nabla^2\phi_1 \\
&&8\pi G_5J^{+z(2)}=-2z^4\cdot\frac{1}{2}\int d\xp\del_z\phi_1\nabla^2\phi_2
\ee

While (\ref{ads5}) has the same functional form as (\ref{ads3}), they are
different in the Laplacian operator. We can check (\ref{ads5}) is again
conserved and the Christoffels involving the
additional directions are accounted for the difference in the Laplacian operators.

With some care, we can obtain the NLO contravariant source, which include
contribution from both LO and NLO covariant sources. The result is shown
in (\ref{J2}) in the main text.

\section{The bulk to boundary propagator}

In this appendix, we want to build a propagator, which produces the stress 
tensor on the boundary field theory when convoluted with the bulk source.
We start with a bulk to bulk propagator for massive scalar defined as follows:

\be\label{bb}
\frac{1}{\sqrt{-g}}\del_\mu(\sqrt{-g}g^{\mu\nu}\del_\nu)G-m^2G=\frac{1}{\sqrt{-g}}\dlt^{(d+1)}(x-x')
\ee

The metric is the Poincare patch of $AdS_{d+1}$. Using the 
Fourier transform: $\tilde{G}(\omg,k,z)=\int G(t,x,z)e^{-i\omg (t-t')+i\vk(\vx-\vx')}$,
(\ref{bb}) takes the following explicit form:

\be\label{FT}
z^2(\omg^2-k^2)\tilde{G}+z^2\del_z^2\tilde{G}+(1-d)z\del_z\tilde{G}=m^2\tilde{G}
=z'{}^{d+1}\dlt(z-z')
\ee

The boundary condition to impose is that $\tilde{G}\rightarrow0$ as
$z\rightarrow0$ and $\tilde{G}$ is outgoing as $z\rightarrow\infty$. 
The solution to (\ref{FT}) is found to be

\be\label{FF_sol}
\tilde{G}=-(zz')^{\frac{d}{2}}I_{\Delta}(\sqrt{k^2-\omg^2}z_<)K_{\Delta}(\sqrt{k^2-\omg^2}z_>)
\ee

where $\Delta=\frac{\sqrt{d^2+4m^2}}{2}$ and $z_>=\max\{z,z'\},\,z_<=\min\{z,z'\}$.

The inverse Fourier transform gives the bulk to bulk propagator:

\be
G(t,x,z)=-\frac{(zz')^{\frac{d}{2}}}{(2\pi)^d}\int I_{\Delta}(\sqrt{k^2-\omg^2}z_<)K_{\Delta}(\sqrt{k^2-\omg^2}z_>)e^{i\omg(t-t')-i\vk(\vx-\vx')}d\omg d^{d-1}k
\ee

Note there are two branch cuts on the real axis $(-\infty,-k)$ and $(k,\infty)$.
The retarded propagator can be obtained if we take the integration contour of
$\omg$ slightly below the real axis: $\omg\rightarrow\omg-i\eps$. We can 
push the integration contour to wrap around the two branch cuts, so that all
the contributions come from two sides of the branch cuts.

\be\label{branch}
&&G(t,x,z,z')=-\frac{(zz')^{\frac{d}{2}}}{(2\pi)^d}\theta(t-t')(\int_{-\infty}^{-k}d\omg+\int_k^{\infty}d\omg)\int d^{d-1}ke^{i\omg(t-t')-i\vk(\vx-\vx')}\times \no
&&\bigg[K_{\Delta}(-i\sqrt{k^2-\omg^2}z_>)I_{\Delta}(-i\sqrt{\omg^2-k^2}z_<)-K_{\Delta}(i\sqrt{k^2-\omg^2}z_>)I_{\Delta}(i\sqrt{\omg^2-k^2}z_<))\bigg] \no
&&=-\frac{(zz')^{\frac{d}{2}}}{(2\pi)^d}\theta(t-t')\int_k^\infty d\omg\int d^{d-1}k 2\pi J_{\Delta}
(\sqrt{\omg^2-k^2}z_>)J_{\Delta}(\sqrt{\omg^2-k^2}z_<)\times \no
&&\sin\omg(t-t')e^{-i\vk(\vx-\vx')}
\ee

Doing the angular integration for the spatial momentum $k$, we obtain:

\be
&&G(t,x,z,z')=-\frac{(zz')^{\frac{d}{2}}}{(2\pi)^{d-1}}\theta(t-t')\int_k^\infty d\omg\int k^{d-2}dkd\Omega^{d-2}\sin\omg(t-t')e^{ikr\cos\theta}\times \no
&&J_{\Delta}(\sqrt{\omg^2-k^2}z_>)J_{\Delta}(\sqrt{\omg^2-k^2}z_<) \no
&&=-\frac{(zz')^{\frac{d}{2}}}{(2\pi)^{d-1}}\theta(t-t')\int_k^\infty d\omg\int k^{d-2}dkd\theta(\sin\theta)^{d-3}d\Omega^{d-2}\sin\omg(t-t')e^{ikr\cos\theta}\times \no
&&J_{\Delta}(\sqrt{\omg^2-k^2}z_>)J_{\Delta}(\sqrt{\omg^2-k^2}z_<) \no
&&=-\frac{(zz')^{\frac{d}{2}}}{(2\pi)^{d-1}}\frac{2^{\frac{d-1}{2}}\pi^{\frac{d-1}{2}}}{r^\frac{d-3}{2}}\theta(t-t'-r)\int_k^\infty d\omg\int dk k^{\frac{d-1}{2}}J_{\frac{d-3}{2}}(kr)\sin\omg(t-t') \times\no
&&J_{\Delta}(\sqrt{\omg^2-k^2}z_>)J_{\Delta}(\sqrt{\omg^2-k^2}z_<)
\ee

We have defined $r=|x-x'|$.
Writing $\beta=\sqrt{\omg^2-k^2}$ allows us to do the $k$-integral\cite{table}:

\be\label{beta}
&&G(t,x,z,z')=-\frac{(zz')^{\frac{d}{2}}}{(2\pi)^{d-1}}\frac{2^{\frac{d-1}{2}}\pi^{\frac{d-1}{2}}}{r^\frac{d-3}{2}}\theta(t-t')\int \frac{\sin\sqrt{\beta^2+k^2}(t-t')}{\sqrt{\beta^2+k^2}}\beta d\beta k^{\frac{d-1}{2}}dk\times \no
&&J_{\Delta}(\beta z_>)J_{\Delta}(\beta z_<)J_{\frac{d-3}{2}}(kr) \no
&&=-\frac{(zz')^{\frac{d}{2}}}{(2\pi)^{d-1}}2^{\frac{d-2}{2}}\pi^{\frac{d}{2}}\theta(t-t'-r)\int J_{\Delta}(\beta z_>)J_{\Delta}(\beta z_<)\beta^\frac{d}{2}J_{-\frac{d-2}{2}}(\beta w)w^{-\frac{d-2}{2}}
\ee

where $w=\sqrt{(t-t')^2-r^2}$.
The final integration of $\beta$ can also be done\cite{table}, we end up with

\begin{align}\label{non-ana}
&G(t-t',x-x',z,z')=-\frac{(zz')^{\frac{d}{2}}}{(2\pi)^{d-1}}2^{\frac{d-2}{2}}\pi^{\frac{d}{2}}\theta(t-t'-r)\times \no
&\left\{\begin{array}{l@{}r}
\sqrt{\frac{2}{\pi^3}}(zz')^{-\frac{d}{2}}(\sinh u)^{-\frac{d-1}{2}}\sin[(-\frac{d}{2}+1-\Delta)\pi]e^{-i\frac{d-1}{2}\pi}Q^{\frac{d-1}{2}}_{\Delta-\frac{1}{2}}(\cosh u)& w>z_>+z_<\\
\frac{1}{\sqrt{2\pi}}(zz')^{-\frac{d}{2}}(\sin v)^{-\frac{d-1}{2}}P^{\frac{d-1}{2}}_{\Delta-\frac{1}{2}}(\cos v)& z_>-z_< <w<z_>+z_<\\
0& \text{otherwise}
\end{array}
\right.
\end{align}

(\ref{non-ana}) is in agreement with early results on bulk to bulk propagator
\cite{DKV,taliotis}. 
However there is a non-analyticity at $w=z_>+z_<$, which
is hidden in (\ref{non-ana}). Integration across the non-analyticity can lead
to finite contribution, thus we choose to start with (\ref{beta}) in building
the bulk to boundary propagator.

The relevant Green's function $G^b(t-t',x-x',z,z')$ is given by:

\be
\frac{z^2}{2}(-\del_t^2+\del_x^2+\del_z^2)G^b+\frac{z}{2}\del_zG^b-4G^b=\dlt(z-z')\dlt^{d}(x-x')
\ee

The metric perturbation in the axial gauge $h_{mn}$ is related to the reshuffled
source $s_{mn}$ by:

\be
h(t,x,z)=\int dz'dt'd^3x's(t',x',z')G^b(t-t',x-x',z,z')
\ee

We have suppressed the tensor indices in $h_{mn}$ and $s_{mn}$. $G^b$ is related
to the bulk to bulk propagator by:

\be
G^b=\frac{2}{z^2z'{}^3}G\vert_{\Delta=2,d=4}
\ee

Let us suppose the source adopts the following expansion near the boundary.

\be
s(t',x',z')=\sum_n s_n(t',x')z'{}^n
\ee

We can perform the integrations first in $z'$ and then in $\beta$ to obtain:

\be
h(t,x,z)=-\frac{1}{2\pi}\int dt'd^3x'\sum_n w^{n-6}z^2\frac{n(n-2)(n-4)}{8}F(\frac{1-n}{2},\frac{3-n}{2};3;\frac{z^2}{w^2})s_n(t',x')
\ee

We are interested in the coefficient of $z^2$, which encodes the boundary stress
tensor. Note $\lim_{z\rightarrow0}F(\frac{1-n}{2},\frac{3-n}{2};3;\frac{z^2}{w^2})\rightarrow1$. The coefficient is given by:

\be
&&-\frac{1}{2\pi}\int dt'd^3x'\sum_n w^{n-6}\frac{n(n-2)(n-4)}{8}s_n(t',x') \no
&&=-\frac{1}{2\pi}\int dt'd^3x'\sum_n w^{n-6}\frac{n(n-2)(n-4)}{8}\frac{1}{n!}\del_z'^n s(t',x',z')\vert_{z'=0} \no
&&=-\frac{1}{2\pi}\int dt'd^3x'dz'\sum_n w^{n-6}\frac{n(n-2)(n-4)}{8}\frac{(-1)^n}{n!} s(t',x',z')\dlt^{(n)}(z')
\ee

We can sum the $n$-series and obtain as our bulk to boundary propagator

\be
&&P^R=-\frac{\theta(t-t'-|x-x'|)}{2\pi}\sum_n w^{n-6}\frac{n(n-2)(n-4)}{8}\frac{(-1)^n}{n!}\dlt^{(n)}(z') \no
&&=-\frac{\theta(t-t'-|x-x'|)}{2\pi}\sum_n w^{n-6}\frac{n(n-1)(n-2)-3n(n-1)+3n}{8}\frac{(-1)^n}{n!}\dlt^{(n)}(z') \no
&&=\frac{\theta(t-t'-|x-x'|)}{2\pi}\bigg[\frac{w^{-3}}{8}\dlt'''(z'-w)+\frac{3w^{-4}}{8}\dlt''(z'-w)+\frac{3w^{-5}}{8}\dlt'(z'-w)\bigg]
\ee

We can further use the property of delta function to replace $\theta(t-t'-|x-x'|)$ by $\theta(t-t')$:

\be\label{PR}
P^R=\frac{\theta(t-t')}{2\pi}\bigg[-\frac{w^{-3}}{8}\dlt'''(z'-w)-\frac{3w^{-4}}{8}\dlt''(z'-w)-\frac{3w^{-5}}{8}\dlt'(z'-w)\bigg]
\ee

\vskip 1cm
%\newpage

\end{document}